\documentclass[english,american,aps,prr,superscriptaddress,reprint]{revtex4-2}
\usepackage[utf8]{inputenc}
\usepackage{xcolor}
\usepackage{babel}
\usepackage{mathtools}
\usepackage{amsmath}
\usepackage{graphicx}
\usepackage{dsfont}
\usepackage[unicode=true,
 bookmarks=true,bookmarksnumbered=false,bookmarksopen=false,
 breaklinks=false,pdfborder={0 0 1},backref=false,colorlinks=true]
 {hyperref}
\hypersetup{
 pdfborderstyle=,pdfborderstyle={},allcolors=blue}

\makeatletter

\newcommand{\mypm}{\mathbin{\mathpalette\@mypm\relax}}
\newcommand{\@mypm}[2]{\ooalign{%
  \raisebox{.1\height}{$#1+$}\cr
  \smash{\raisebox{-.6\height}{$#1-$}}\cr}}

\usepackage{babel}

\usepackage{bbold}

\usepackage{algpseudocode}

\makeatother

\begin{document}
\title{
Correlation-boosted quantum engine: A proof-of-principle demonstration}

\author{Marcela Herrera}
\email{Electronic Address: amherrera@uao.edu.co}

\affiliation{Centre for Bioinformatics and Photonics—CIBioFi, Edificio E20
No.~1069, Universidad del Valle, 760032 Cali, Colombia}
\affiliation{Departamento de Física, Universidad del Valle, 760032 Cali, Colombia}
\affiliation{Departamento de Ciencias Naturales, Universidad Autónoma de Occidente,
Cali, Colombia}

\author{John H. Reina}
\email{Electronic Address: john.reina@correounivalle.edu.co}
\affiliation{Centre for Bioinformatics and Photonics—CIBioFi, Edificio E20
No.~1069, Universidad del Valle, 760032 Cali, Colombia}
\affiliation{Departamento de Física, Universidad del Valle, 760032 Cali, Colombia}
\author{Irene D'Amico}
\email{Electronic Address: irene.damico@york.ac.uk }

\affiliation{Department of Physics, University of York, York YO10 5DD, United Kingdom}

\author{Roberto M. Serra}
\email{Electronic Address: serra@ufabc.edu.br}

\affiliation{Centro de Ciências Naturais e Humanas, Universidade Federal do ABC,
Avenida dos Estados 5001, 09210-580 Santo André, São Paulo, Brazil}

\affiliation{Department of Physics, Zhejiang Normal University, Jinhua 321004, China}

\begin{abstract}
Employing currently available quantum technology, we design and implement a non-classically correlated SWAP heat engine that allows to achieve an efficiency above the standard Carnot limit. Such an engine also 
boosts the amount of extractable work, in a wider parameter window,
with respect to engine's cycle in the absence of initial quantum correlations in the working substance.
The boosted 
efficiency arises from a trade-off between  the entropy production and the consumption of quantum correlations
during the full thermodynamic cycle.
We derive a
generalized second-law limit for the correlated cycle
and implement a proof-of-principle  demonstration of the engine efficiency enhancement by effectively tailoring the thermal engine on a cloud quantum processor.

\end{abstract}
\maketitle

\section{Introduction}
Recent years have witnessed the rise of quantum thermodynamics (QTD),
which has rapidly become  an arena to test and debate fundamental
concepts such as the laws of thermodynamics and the related
fluctuation theorems at the quantum scale~\citep{Adesso2018,Deffner2019,Kosloff2013,Goold,Anders2016,Brandao2015,Jarzynski1997,Crooks1999,Esposito2009,Campisi2011,Herrera2021,Auffeves2022}. In parallel, QTD is facilitating the extension
of practical concepts such as heat and work towards the
design~\citep{Hugel2002,Steeneken2011,Blickle2012,Brantut2013,Lutz2014,Thierschmann2015,Rossnagel2016,Schmidt2018} and, more recently, the implementation~\citep{Deffner2019,Solfanelli2021,Zou2017,Assis2019,Klatzow2019,Peterson2019, Denzler2021, Myers2022}  of thermal machines based on a handful of quantum degrees of freedom. Particular attention is being
given to the concept of entropy production for systems out of
equilibrium~\cite{
Labon2008,Batalhao2014,Batalhao2015,Camati2016,Landi2021} and to the -- related -- second law of
thermodynamics~\cite{Brandao2015,Seifert2012,Martinez2015}, as well as to the role of quantum correlations in both fundamental concepts~\cite{Bera2017,jh2022,Sapienza2019,jh2017,Zawadzki2020} and in the functioning of thermal
devices~\cite{Henao2018,Micadei2019,Revathy2020,Watanabe2020,Mukherjee2021,Asadian2022}.
Quantum features can be considered as extra resources. Squeezed states, for instance, can  enhance  the performance of a microscopic engine above its classical limits~\cite{Lutz2014,OliveiraJ2022}. Coherence with a dynamical interference~\cite{Francica2019, Santos2019,Camati2019a}, quantum measurements~\cite{Alonso2016,Ding2018,Bresque2021,Buffoni2019,Elouard2018,Anka2021,Lisboa2022}, and quantum  operations causal order ~\cite{Rubino2021,Felce2020,Cao2021,Felce2021,Simonov2022,Dieguez2022} also play nontrivial roles in the performance of thermodynamic tasks~\cite{Henao2018,Micadei2019}. 

Through the design and proof-of-principle  implementation of a
two-qubit thermal machine (on a quantum processor), here, we demonstrate that the use of quantum correlations as an extra resource can indeed lead to a  generalized second law of thermodynamics encompassing regimes with efficiency larger than the standard Carnot limit. As an added bonus, quantum correlations increase the amount of extractable work, as well as extend the
parameter region corresponding to useful work extraction for  the proposed cycle.

This paper is organized as follows. 
Taking advantage of initial non-classical correlations in a two-qubit working substance, in Sec.~\ref{qdesign} we design a correlated quantum heat-engine based on the concept of a partial SWAP operation. 
The engine's nonequilibrium-thermodynamics quantifiers are introduced and analytically computed in Sec.~\ref{WQ}.
In Sec.~\ref{qesetup}, we introduce a quantum processor implementation of the  quantum heat engine setup and experimentally  demonstrate the performance enhancement of the proposed correlated SWAP heat engine.
In Sec.~\ref{booster},
we derive an 
analytical expression for the SWAP engine efficiency and provide a  criterion for performance over the classical limit (above Carnot) to occur. In Sec.~\ref{eff},
we prove that such an engine can exceed the conventional classical limits,
with an out-performance that is well described by an information-to-energy trade-off relation for the cycle efficiency, which can be seen as a  quantum generalized efficiency limit for a two-stroke cycle in the presence of non-classical correlations.
In Sec.~\ref{exp} and~\ref{tomo},
based on the ancilla-assisted two-point-measurement method developed in Ref.~\citep{Solfanelli2021},  we use a cloud quantum processor~\citep{ibmq} 
to implement an experimental proof-of-principle of our design.
Summary of results and concluding remarks are given in Sec.~\ref{conc}.

\section{Design and functioning of a correlated quantum heat engine}
\label{qdesign}

\begin{figure}[ht]
\begin{centering}
\includegraphics[scale=0.39]{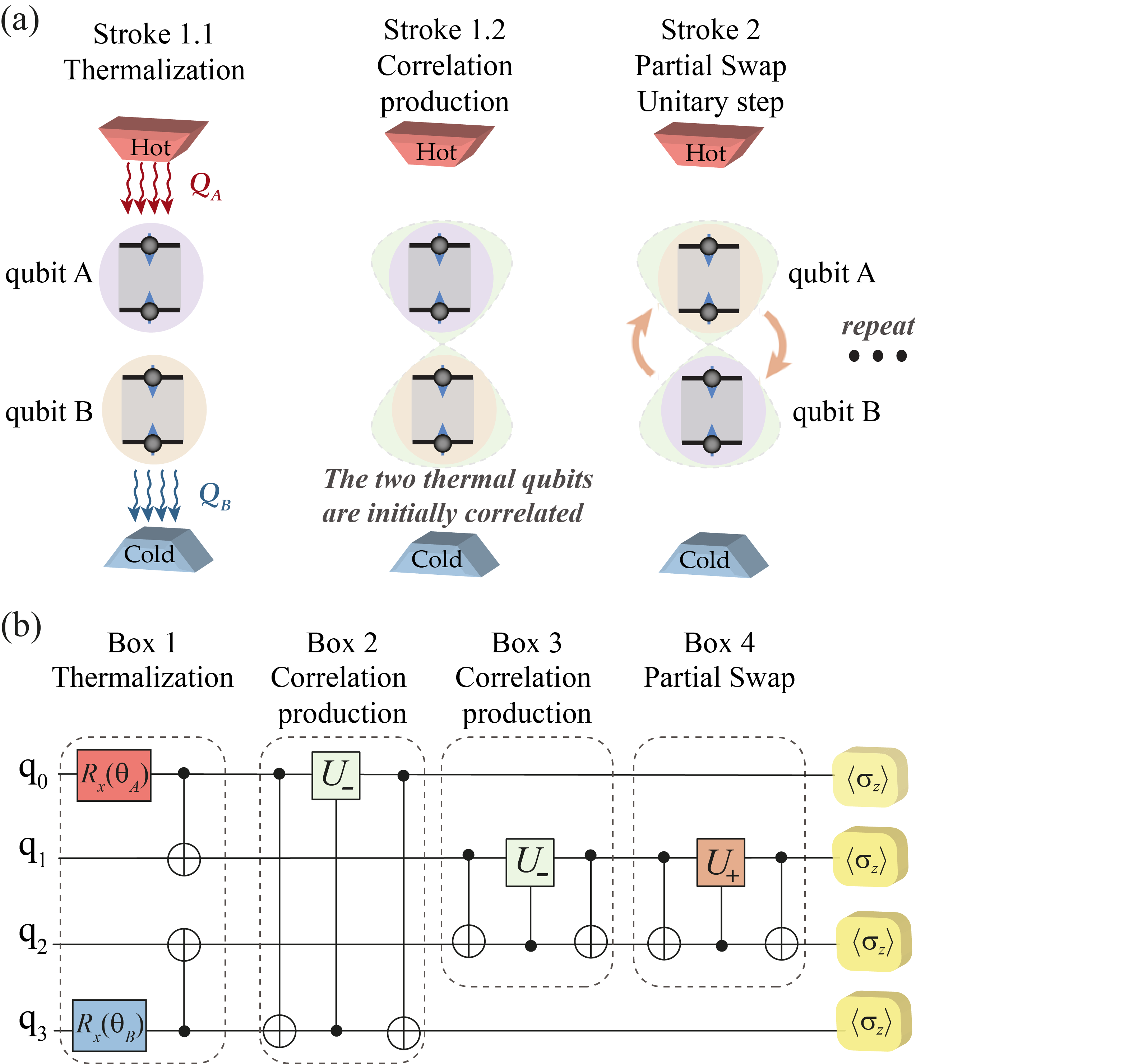}
\par
\end{centering}
\caption{(a) 
Thermodynamic cycle of a correlated SWAP quantum heat engine. The two-qubit ($AB$) system acts as the working substance. 
$Q_A$ ($Q_B$) denotes the heat exchanged with the hot (cold) environment.
(b) Quantum-circuit to test the correlated heat engine concept. 
The input to the circuit is the ground state $\left|0000\right\rangle$. 
The qubits $q_{0}$ and
$q_{3}$ are employed as ancillae to prepare the engine in the initial thermal
state $\rho_{1}^{0}\otimes\rho_{2}^{0}\equiv\tilde{\rho}_{A}^{0}\otimes\tilde{\rho}_{B}^{0}$
(box 1); the rotation angles $\theta_i$ depend on $\alpha$ and the qubits effective temperature. The $AB$ correlations are  generated in boxes 2 and 3 via $U_-$, and the final partial SWAP implementation, via $U_+$, in  box 4.
}
\label{figcycle}
\end{figure}

The quantum engine is composed by two qubits with energy gaps $\varepsilon_{A}$
and $\varepsilon_{B}$. Each of them is initially coupled to an effective
heat environment, as schematized in Fig.~\ref{figcycle}(a), stroke 1.1, which lead to the product of Gibbs states,  $\tilde{\rho}_{A}^{0}\otimes\tilde{\rho}_{B}^{0}$. After such a complete thermalization, the two 
qubits get correlated through stroke 1.2 (Fig.~\ref{figcycle}(a)) in a state given by
\begin{equation}
\rho_{AB}^{0}=\rho_{A}^{0}\otimes\rho_{B}^{0}+\chi_{AB},
\label{inistate}
\end{equation}
where the reduced local state for each qubit $\rho_{i}^{0}=\text{exp}\left(-\beta_{i}\mathcal{H}_{i}\right)/\mathcal{Z}_{i}$
is a thermal equilibrium state at the inverse temperature  $\beta_{i}=\left(k_{B}T_{i}\right)^{-1}$, $i=A,B$;
$\mathcal{Z}_i=
\text{Tr}_i\, \text{exp}\left(-\beta_{i}\mathcal{H}_{i}\right)$ is the corresponding partition function, and $\chi_{AB}=\alpha\left|01\right\rangle \left\langle 10\right|+\alpha^*\left|10\right\rangle \left\langle 01\right|$
gives the relevant correlation term, 
with $\text{Tr}_i\chi_{AB}=0$. Here, the states $\left|0\right\rangle$ and $\left|1\right\rangle$ denote the ground and excited eigenstates of the qubit Hamiltonian 
$\mathcal{H}_{i}=-\frac{1}{2}\varepsilon_i \sigma_Z^{(i)}$; the total two-qubit 
Hamiltonian
$\mathcal{H}_{AB}=\mathcal{H}_{A}+\mathcal{H}_{B}$. 
Qubit $A$ is assumed
to be hotter than the qubit $B$, that is $\beta_{A}<\beta_{B}$.  We remark that the correlation term $\chi_{AB}$ does not contribute to the energy of qubits $A$ and $B$. The populations of states $\tilde{\rho}_{A}^{0}$ and $\tilde{\rho}_{B}^{0}$ are chosen in a way to obtain the desired correlated state in Eq.~(\ref{inistate}) (see Appendix~\ref{energyexchange}).
The energy exchange  between the two
qubits is determined by means of a partial SWAP operation, as illustrated
in Fig~\ref{figcycle}(a), stroke 2, where work can be extracted from or performed on the quantum system.  Physically, this can be implemented  by
an effective unitary Heisenberg exchange Hamiltonian evolution ($j=X,Y,Z$)
\begin{equation}
\mathcal{U}_t= \exp\left(\frac{-i  J t}{2}\sum_{j} \sigma^{(A)}_j\otimes\sigma^{(B)}_j\right).
\label{ham}
\end{equation}
At $t=0$  there is no interaction at all;  a complete  SWAP operation~\cite{Campisi2015,Timpanaro2019}  takes place at $t=\pi /(2J)$.

Although the initial state Eq.~(\ref{inistate}) seems to be similar to the one considered in Ref.~\cite{Micadei2019}, we stress that  the off-diagonal parameter has a different phase and the evolution is driven by a different Hamiltonian, which makes all the difference to exploit non-convectional energy  flows to thermal tasks. Moreover, the correlation parameter used in~\cite{Micadei2019} would not allow for an advantage in the quantum heat engine here implemented.

Next, we consider a proof-of-principle implementation on a quantum processor. Figure~\ref{figcycle}(b) shows a quantum circuit that implements a partial SWAP engine in the presence of initial correlations. 
It is based on the use of the (controlled) gate $U_{\mypm}(x)\equiv\mathds{1}\sqrt{1-x}\pm i\sigma_{Y}\sqrt{x}$, as follows. Qubits $q_{1}$ and
$q_{2}$ constitute the working substance, while qubits $q_{0}$ and
$q_{3}$ are employed as ancillae to prepare the engine in the initial
thermal state $\rho_{1}^{0}\otimes\rho_{2}^{0}$,
and emulate the actions of the hot and cold environments, respectively. This initial state is prepared by means of a rotation gate $R_x(\theta_i)$ in the ancillae and a CNOT gate between the ancillae and the qubits $i=A,B$ 
($q_1,q_2$ in box 1,  Fig.~\ref{figcycle}(b)). The rotation angles are associated with the effective temperature of each qubit and with the correlation parameter $\alpha$ as follows: 
\begin{eqnarray}
\theta_A &=&\arccos{\sqrt{p_-}},
\; \; \;
\theta_B = \arccos{\sqrt{p_+}},
\label{eq:ang1}
\\
p_{\pm} &=& \frac{1}{2}\left(p_{A}+p_{B}\pm\sqrt{(p_{B}-p_{A})^{2}+4\alpha^{2}}\right),
\label{eq:ang2}
\\ 
\frac{\varepsilon_B}{\varepsilon_A} &=& \frac{\beta_A \ln\left( p_B\mathcal{Z}_B\right)}{\beta_B \ln\left( p_A\mathcal{Z}_A\right)}
\label{eq:ang3},
\end{eqnarray}
 where $p_i=\exp(-\beta_i\varepsilon_i)/\mathcal{Z}_{i}$ and $\varepsilon_B/\varepsilon_A$ gives  the qubits energy-gap ratio. 

In the present implementation, the initial correlation in the working substance ($q_1$ and $q_2$) is achieved by a CNOT gate framing one   controlled $U_- (x)$ gate,  $x=x(\alpha)=(p_A - p_B)/(p_{+}-p_{-}) +1/2$, as shown in Fig.~\ref{figcycle}(b) (box 3) \cite{Note_corr}, and \ensuremath{\alpha}
should satisfy $\alpha\leq1/\left(\mathcal{Z}_{A}\mathcal{Z}_{B}\right)$ to fulfill that the system's density operator is a positive semi-definite. We also produced the same reduced correlated state of the working substance in the ancillary qubits $q_0$ and $q_3$, through box 2 of Fig.~\ref{figcycle}(b), in order to have a copy of the initial correlated state at the circuit end, thus having an abridged measurement strategy.
After boxes 2 and 3 (Fig.~\ref{figcycle}(b)), the working substance is in the state given by  Eq.~(\ref{inistate}).
Two distinct scenarios are identified: i) $\alpha=0$, the qubits are initially uncorrelated, and ii) $\alpha\neq0$, 
the bipartite state is initially  correlated.
The partial SWAP operation, stroke 2, can be  effectively implemented through a quantum circuit
composed by two CNOT gates framing one controlled-gate 
$U_+(\lambda)$ (Fig.~\ref{figcycle}(b), box 4). 
For $\lambda=1$, we have a full SWAP between the qubit states, while for  $\lambda=0$  there is no operation at all between such
qubits.
The partial SWAP takes place at a time $\tau$, with $0<\tau<\pi/(2J)$.
We finally measure all qubits in their respective ($\sigma_{Z}^{(i)}$) energy basis.
\begin{figure}
\begin{centering}
\includegraphics[scale=0.33]{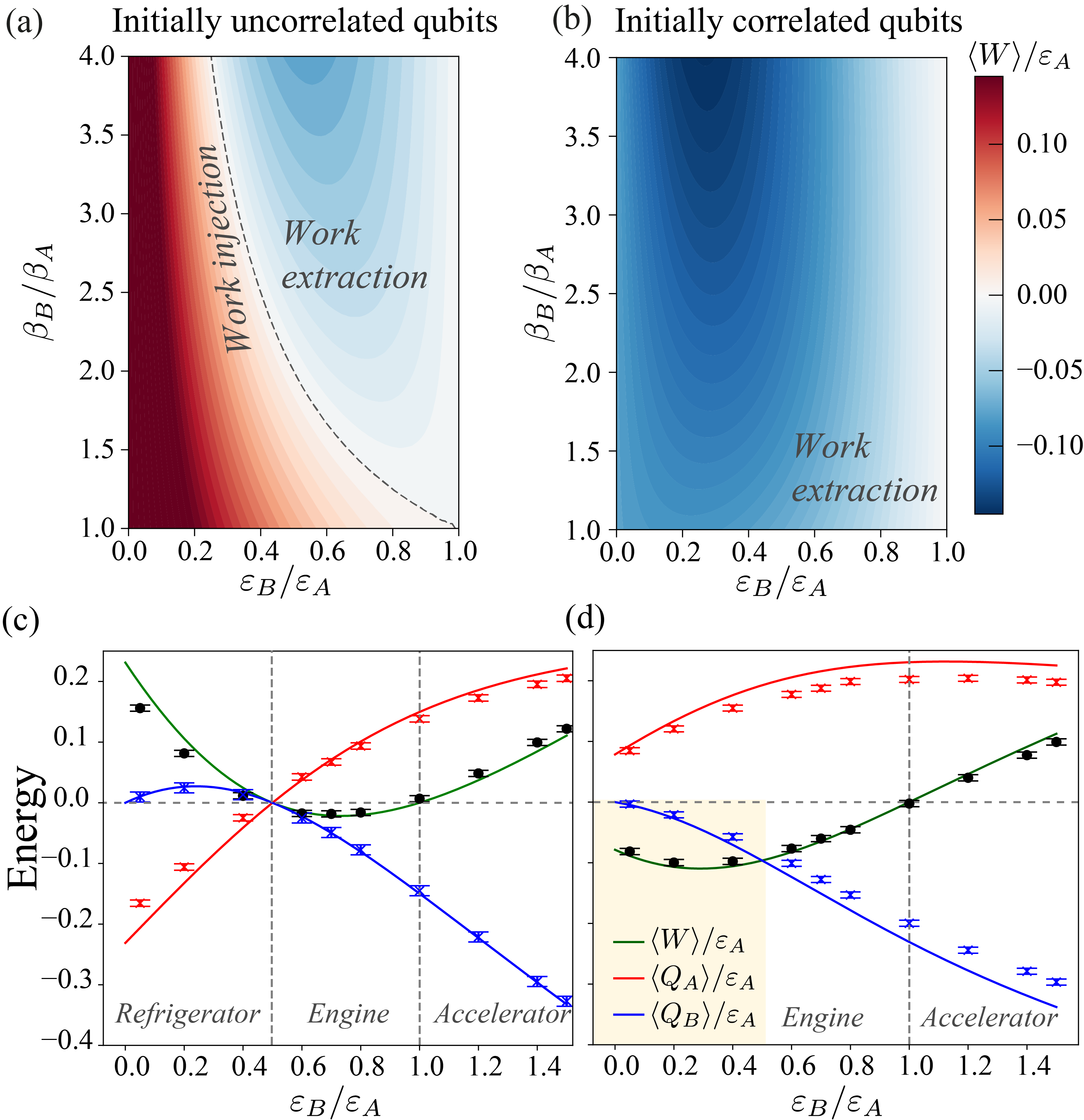}
\par\end{centering}
\caption{Parameters phase diagram for engine operation mode:
(a) initially uncorrelated qubits, work extraction is only allowed for 
parameter  ratios  $\{\beta_{B}/\beta_A, \varepsilon_{B}/\varepsilon_A\}$  above the dashed curve (blue region); (b)  initially correlated qubits, work extraction is allowed for all  $0\leq\varepsilon_B/\varepsilon_A<1$, 
$\beta_{A}<\beta_{B}$.  Experimental results for the  re-scaled average work ($\left\langle W\right\rangle/\varepsilon_A$), and heat from the hot ($\left\langle Q_A\right\rangle/\varepsilon_A$) and cold ($\left\langle Q_B\right\rangle/\varepsilon_A$) reservoirs for (c) initially uncorrelated qubits, and (d) initially correlated qubits; in all experimental runs in the quantum processor, we set: 
$\beta_{B}=2\beta_{A}$,
$\lambda=0.6$,
and 
$\alpha_{max}=1/\left(\mathcal{Z}_{A}\mathcal{Z}_{B}\right)$. The error bars were estimated using the standard deviation of the measured data. The solid curves correspond to the theoretical prediction. 
\label{figenergy}}
\end{figure}
\begin{figure*}
\begin{centering}
\includegraphics[scale=0.35]{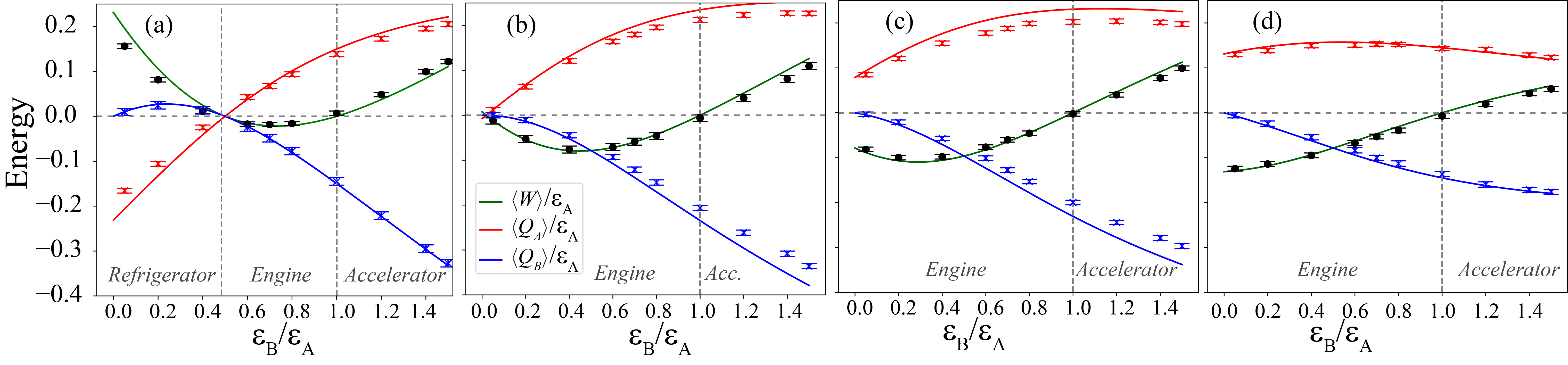}
\par\end{centering}
\caption{
Quantum-processor-implementation results for the  average work, and heat from the hot ($\left\langle Q_{A}\right\rangle$) and cold ($\left\langle Q_{B}\right\rangle$) reservoirs for (a) initially uncorrelated qubits
 ($\lambda=0$), and  initially correlated qubits: (b) $\lambda=0.8$, (c) $\lambda=0.6$, and (d) $\lambda=0.2$. In all runs in the quantum processor, $\beta_{B}=2\beta_{A}$
and  $\alpha_{max}=1/\left(\mathcal{Z}_{A}\mathcal{Z}_{B}\right)$. The error bars were estimated using the standard deviation of the measured data. The solid curves are obtained from our theoretical predictions from Eqs.~\eqref{eq:workeng} to~\eqref{eq:qhot} and numerical simulations.  
\label{fig1_sm}}
\end{figure*}

\section{Quantum-engine nonequilibrium-thermodynamics quantifiers}
\label{WQ}

For  the aforementioned cycle, we calculate the mean energies involved in the whole process: the  average values for the work, $\langle W \rangle=\text{Tr}[(\rho_{AB}^{f}-\rho_{AB}^{0})\mathcal{H}_{AB}]$, (that takes place in stroke 2)
and the 
heat contributions from the hot ($\left\langle Q_{A}\right\rangle$) and cold  ($\left\langle Q_{B}\right\rangle$) environments, $\left\langle Q_{i}\right\rangle=-\text{Tr}[(\rho_{i}^{f}-\rho_{i}^{0})\mathcal{H}_{i}]$, where  
$\rho_{A(B)}^{f}=\text{Tr}_{B(A)}\left(\rho_{AB}^f\right)= \text{Tr}_{B(A)}\left(\mathcal{U}_\tau\rho_{AB}^0 \mathcal{U}^\dagger_\tau\right)$
is the final out-of-equilibrium reduced state for qubit $A$ ($B$) after stroke 2.
We obtain:
\begin{eqnarray}
\left\langle W \right\rangle & = & 2\left(\varepsilon_{B}-\varepsilon_{A}\right) f(\Delta\nu,\lambda,\alpha),
\label{eq:workeng}
   \\
\left\langle Q_A \right\rangle  & = & 2\varepsilon_{A} f(\Delta\nu,\lambda,\alpha) 
\label{eq:qcold},
\\ 
\left\langle Q_B \right\rangle & = & -2\varepsilon_{B} f(\Delta\nu,\lambda,\alpha)
\label{eq:qhot},
\end{eqnarray}
 with
$f(\Delta\nu,\lambda,\alpha)=\frac{\sinh(\ensuremath{\Delta}\ensuremath{\nu})}{\mathcal{Z}_{A}\mathcal{Z}_{B}}\lambda+2\alpha\sqrt{\lambda(1-\lambda)}$, where $\text{\ensuremath{\Delta}\ensuremath{\nu=}\ensuremath{\left(\varepsilon_{B}\beta_{B}-\varepsilon_{A}\beta_{A}\right)}/2}$.  
Since  the total energy is conserved, Eq.~\eqref{eq:workeng} to \eqref{eq:qhot} 
 fulfill energy conservation, i.e.,   $\left\langle W\right\rangle=-(\left\langle Q_A\right\rangle+\left\langle Q_B\right\rangle)$. 
 These results for the correlated case ($\alpha\ne 0$) are plotted in Fig.~\ref{figenergy}(b) and as the continuous curves in Fig.~\ref{figenergy}(d) (see also Fig.~\ref{fig1_sm}). For comparison, in  Fig.~\ref{figenergy}(a) and (c) we plot the corresponding energies in the absence of initial correlations ($\alpha=0$). 

We characterized the performance of the quantum engine by varying the energy gap ratio $\varepsilon_B/\varepsilon_A$, setting $\beta_{B}=2\beta_{A}$, and fixing the SWAP  parameter to $\lambda = 0.6$ (other values are considered in Fig. \ref{fig1_sm}). For the correlated initial state we considered  $\alpha$'s maximum value, $\alpha_{max}=1/\left(\mathcal{Z}_{A}\mathcal{Z}_{B}\right)$.

  In Fig.~\ref{figenergy}(a) and (b) we plot the
 parameters'  diagram 
for  temperature and energy gap ratios
 $\{\beta_{B}/\beta_A, \varepsilon_{B}/\varepsilon_A\}$
required for work extraction  for both, initially    uncorrelated, and correlated scenarios.
The dashed curve in the diagram for the case without initial correlations, Fig.~\ref{figenergy}(a),  separates the regions for which the  system can work as a refrigerator (work injection) or as a heat engine (work extraction). 
In contrast,
for initially correlated qubits (Fig.~\ref{figenergy}(b)), work injection gets suppressed in the whole gap ratio $\varepsilon_B/\varepsilon_A< 1$, for all $\beta_{A}<\beta_{B}$, and work extraction becomes much higher; the maximal extracted work can be seen, e.g.,  for  $\beta_B\gtrsim 3 \beta_A$ and $0.2\lesssim \varepsilon_B/\varepsilon_A\lesssim 0.5$ (Fig.~\ref{figenergy}(b)).

In Fig.~\ref{figenergy}(c) and (d) we plot the mean energy (re-scaled) quantities, work ($\left\langle W\right\rangle$, black dots) and heat from the hot ($\left\langle Q_A\right\rangle$, red dots) and cold ($\left\langle Q_B\right\rangle$, blue dots) environments obtained from the experimental runs in the quantum processor. The error bars were estimated using the standard deviation of the measured data in the quantum processor. 
Three operational regimes~\cite{Timpanaro2019} appear in the initially uncorrelated qubits scenario  (Fig.~\ref{figenergy}(c)):
refrigerator 
($0<\varepsilon_B/\varepsilon_A<1/2$), heat engine
($1/2<\varepsilon_B/\varepsilon_A<1$), and 
heat accelerator
($\varepsilon_B/\varepsilon_A>1$). However,
when the qubits are initially correlated, 
the partial SWAP engine only  exhibits two operational modes: heat engine ($0<\varepsilon_B/\varepsilon_A<1$) and 
heat accelerator
($\varepsilon_B/\varepsilon_A>1$) (Fig.~\ref{figenergy}(d)).  In the engine operation mode, quantum correlations boost the amount of work that can be extracted, making it at its maximum about an order of magnitude larger than the one obtained in the absence of initial qubit correlations. As already explained above, this result has also been  verified for other $\lambda$ values (see Fig.~\ref{fig1_sm}). In Figs.~\ref{figenergy}(c), (d) 
and Fig.~\ref{fig1_sm} we find a very good agreement between the quantum-processor-implementation results and the corresponding theoretical prediction from Eqs.~(\ref{eq:workeng}) to (\ref{eq:qhot}) 
(solid curves). 
Furthermore, quantum correlations enlarge the  $\varepsilon_B/\varepsilon_A$ values' window where work extraction is possible.

In Fig.~\ref{fig1_sm}  the average work and heat from the hot and cold environments, for different $\lambda$ values are plotted. As established above, Fig.~\ref{fig1_sm}(a) shows the three different modes of operation of the SWAP quantum engine that are available if we consider initially  uncorrelated qubits ($\lambda=0$). The engine mode occurs for $0.5<\varepsilon_B/\varepsilon_A<1$. Otherwise, for initially  correlated qubits, the engine mode expands its range to $0.0<\varepsilon_B/\varepsilon_A<1$, as it is shown in Fig.~\ref{fig1_sm}(b)-(d),   $\lambda=0.8,0.6,0.2$, respectively. The lower the  $\lambda$ parameter the greater the amount of extracted work. In fact, executing a full SWAP ($\lambda=1$) in the engine
would erase the advantage due to its quantum correlations hence obtaining the same result as in the absence of initial correlations.

From Eqs.~\eqref{eq:workeng} and \eqref{eq:qcold} it is straightforward to obtain the SWAP engine efficiency, 
\begin{equation}
   \eta=-\frac{\left\langle W\right\rangle }{ \left\langle Q_A\right\rangle} = 1-\frac{\varepsilon_{B}}{\varepsilon_{A}}.  \label{eff}
 \end{equation}
For  qubit energies such that    $\frac{\varepsilon_{B}}{\varepsilon_{A}} = \frac{\beta_{A}}{\beta_{B}}$, the  quantum engine achieves the standard Carnot limit, 
\begin{equation}
\eta_{Carnot}\equiv 1-\frac{\text{\ensuremath{\beta_{A}}}}{\beta_{B}}  .
  \label{carnot}
\end{equation}
\begin{figure*}
\begin{centering}
\includegraphics[scale=0.47]{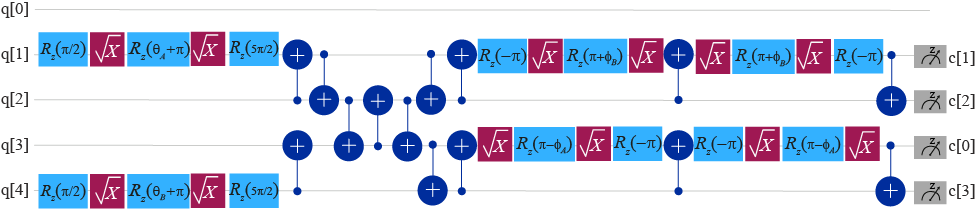}
\par\end{centering}
\caption{
Schematic representation of the transpiled circuit implemented by 3  basis gates: $R_z$ rotations, $\sqrt{X}$, and CNOT. The circuit depth or “layers” of quantum gates are  executed in parallel according to \textsl{ibmq\_manila} topology to achieve the computation indicated in the original  circuit of  Fig.~1(b).
The $q[i]$ input labeling indicates the physical qubits used by the 5-qubit processor and the circuit output $c[i]$ represents the classical data where the measurement result is recorded.}
\label{fig_transpiled}
\end{figure*}

\section{QUANTUM HEAT ENGINE EXPERIMENTAL SETUP}
\label{qesetup}

Here we give further details about the quantum processor implementation  of a quantum heat engine and corresponding  data analysis.
For the implementation and characterization of the quantum heat engine we ran several experiments on the 5-qubit \textsl{ibmq\_manila} quantum processor~\citep{ibmq}. Below we experimentally  demonstrate the performance enhancement of a correlated SWAP heat engine and confirm
our  theoretical predictions.

For each run in the quantum processor, we implemented the quantum  circuit depicted in Fig.~1(b) and collected the qubits statistics over a sample of size 20000. We performed the experiment for different values of the partial SWAP parameter $\lambda$. Additionally, for each  $\lambda$, we ran 10 experiments with the aforementioned sample size in order to see the fluctuations of the experimental setup. From these, we estimated the error propagation, using the standard deviation. Each circuit run is determined by the correlation  ($\alpha$) and thermalization ($\lambda$) parameters. 

\noindent 
\begin{table}[h]
\centering{}%
\begin{tabular}{ccccccc}
\hline \hline 
  \multicolumn{7}{c}{\textbf{Engine implementation experimental parameters}}
   \tabularnewline
 &  & & & $\lambda=0.2$& $\lambda=0.6$& $\lambda=0.8$\tabularnewline
 $\varepsilon_{B}/\varepsilon_{A}$ & $\theta_A(\text{rad})$ & $\theta_B(\text{rad})$ & $\phi_A(\text{rad})$& $\phi_B(\text{rad})$& $\phi_B(\text{rad})$& $\phi_B(\text{rad})$\tabularnewline
\hline 
\hline 
0.05 & 2.41  & 1.34 & 2.57 & 0.10 &-0.32 & -0.54\tabularnewline
\hline 
0.20 & 2.47  & 1.45 & 2.50 & 0.17 &-0.25 & -0.47\tabularnewline
\hline 
0.40 & 2.55  & 1.56 & 2.41 & 0.27 &-0.15 & -0.37\tabularnewline
\hline 
0.60 & 2.64  & 1.69 & 2.31 & 0.37 &-0.05 &-0.27\tabularnewline
\hline 
0.70 & 2.67  & 1.74 & 2.26 & 0.42 &-0.001 & -0.22\tabularnewline
\hline 
0.80 & 2.71  & 1.79 & 2.21 & 0.467&0.05 &-0.17\tabularnewline
\hline 
1.00 & 2.78  & 1.86 & 2.12 & 0.56  &0.14 &-0.08\tabularnewline
\hline 
1.20 & 2.84  & 1.92 & 2.03 & 0.65 & 0.22 & 0.002\tabularnewline
\hline 
1.40 & 2.89  & 1.96 & 1.96 & 0.72 &0.30 & 0.08\tabularnewline
\hline 
1.50 & 2.91  & 1.97 & 1.92 & 0.75 &0.33 &0.11\tabularnewline
\hline 
\hline 
\end{tabular}\caption{Parameters used in the transpiled circuit displayed in
Fig.~\ref{fig_transpiled} for the engine implementation in a quantum processor.
}
\label{Tableangles}
\end{table}

In Figure~\ref{fig_transpiled}, we outline the transpiled circuit that has been 
implemented following the topology and optimization of \textsl{ibmq\_manila}. This corresponds to the engine thermodynamic cycle portrayed in the circuit of Fig.~1(b). The correlated SWAP engine implementation  parameters (see Table~\ref{Tableangles}) have 
energy-gap ratio $\varepsilon_B/\varepsilon_A$, and angles
$\theta_A$ and 
$\theta_B$, with $p_{\pm}$ and $p_{i}$ as given in Eqs.~\eqref{eq:ang1} to \eqref{eq:ang3}. Here,
\begin{eqnarray}
   \phi_A &=&\arcsin{\sqrt{x(\alpha)}},\\
   \phi_B &=& \phi_A - \arcsin{\sqrt{\lambda}},\\
   x(\alpha)&=&(p_A - p_B)/(p_{+}-p_{-}) +1/2,
\end{eqnarray} 
where $x(\alpha)$ and $\lambda$ give the correlation and partial swap parameters, respectively. 
In Table~\ref{Tableangles}  we display the experimental angles used in all the runs in the quantum processor here reported.

\noindent 
\begin{table}[h]
\centering{}%
\begin{tabular}{cccc}
\hline \hline 
  \multicolumn{4}{c}{\textbf{\textsl{ibmq\_manila}  calibration data}}\tabularnewline
Qubit & $T_{1}(\mu\text{s})$ & $T_{2}(\mu\text{s})$ & Gate time (ns)\tabularnewline
\hline 
\hline 
$q_{0}$ & 177.13  & 78.73  & 0\_1: 277.33\tabularnewline
\hline 
{$q_{1}$} & {186.02 } & {75.55 } & 1\_2: 469.33\tabularnewline
 &  &  & 1\_0: 312.89\tabularnewline
\hline 
{$q_{2}$} & {136.19 } & {22.30 } & {2\_3: 355.56}\tabularnewline
 &  &  & 2\_1: 504.88\tabularnewline
\hline 
{$q_{3}$} & {184.82 } & {46.64 } & 3\_4: 334.22\tabularnewline
 &  &  & 3\_2: 391.11\tabularnewline
\hline 
$q_{4}$ & 122.91  & 43.53  & 4\_3: 298.67\tabularnewline
\hline 
\hline 
\end{tabular}\caption{Calibration data for the qubits involved in the generation of the heat engine at the 5-qubit \textsl{ibmq\_manila} quantum processor (Fig.~\ref{fig_transpiled}).}
\label{TableS1}
\end{table}

In the transpiled circuit (Fig.~\ref{fig_transpiled}), three basic gates are used; $R_z$ rotations, $\sqrt{X}$ and CNOT, in order to parallel implement the original quantum circuit of  Fig.~1(b). We resort to the gating and decoherence  times 
for \textsl{ibmq\_manila}~\citep{ibmq} 
 as figures of merit to  check that the  coherence properties required in the implementation of the quantum thermodynamic cycle 
are guaranteed. Clearly, relaxion ($T_1$) and decoherence ($T_2$) times are about three orders of magnitude longer than the qubits' gating times. These times are reported in Table~\ref{TableS1}. 

  The last column gives the execution times of the `$i\_j$'
   two-qubit gates according to the \textsl{ibmq\_manila} topology ($i,j=0,1,...,4$). By assuming that each gate takes its longest possible execution time ($\sim 500$~ns),  we can overestimate the runtime of the full circuit to 
 approximately $11\;\mu\text{s}$, which is shorter than the shortest $T_2$ time reported in Table~\ref{TableS1}.
 
\begin{figure}
\begin{centering}
\includegraphics[scale=0.34]{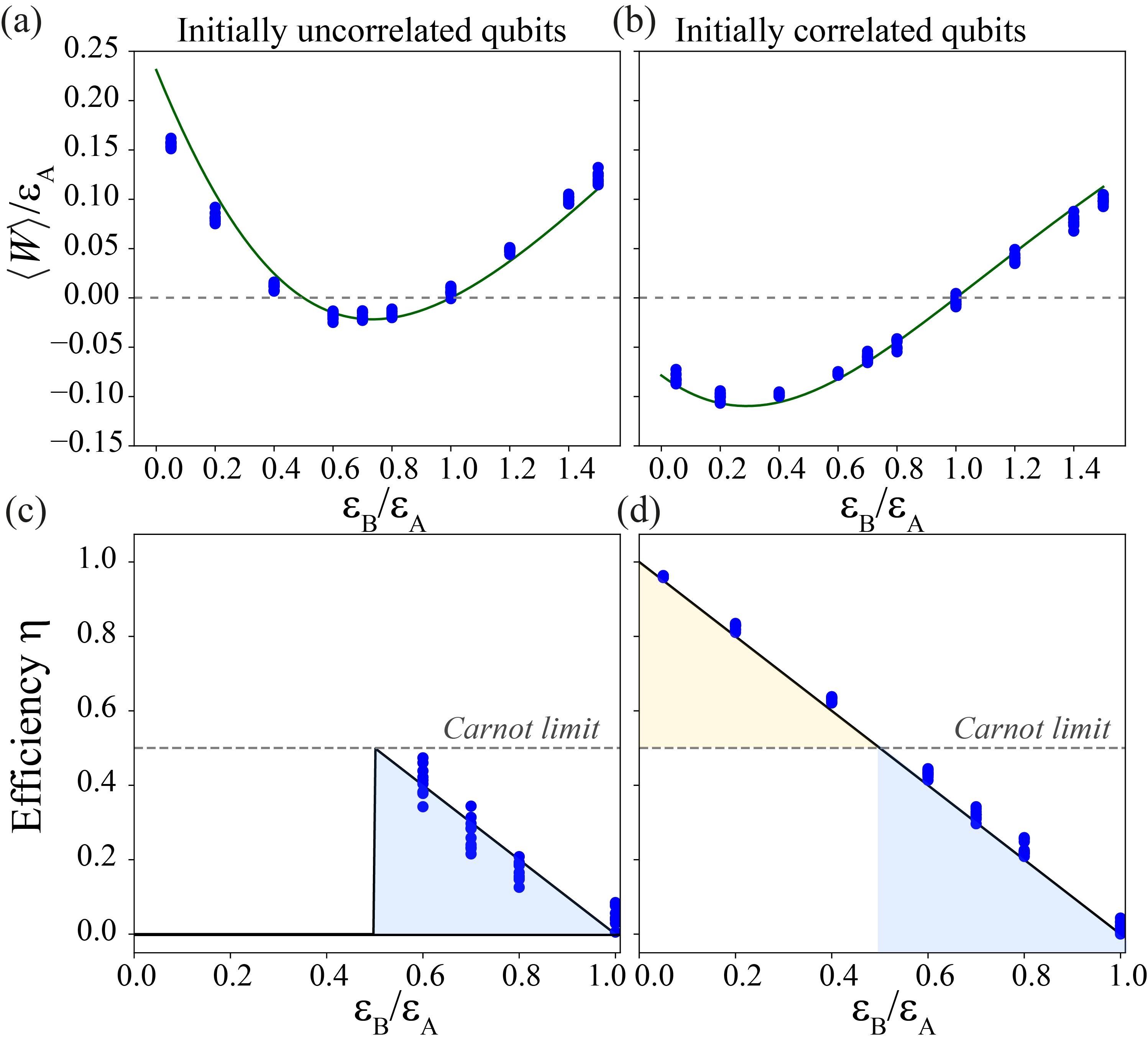}
\par\end{centering}
\caption{
Quantum-processor-implementation results for the average work 
 and efficiency, considering (a) and (c) initially uncorrelated qubits; (b) and (d) initially correlated qubits. The blue dots indicate the realization of 10  different experiments for each $\varepsilon_B/\varepsilon_A$ value, and each one performed with 20000 shots. The full lines in (c) and (d) correspond to the theoretical prediction. In all the figures, $\lambda=0.6$, $\beta_{B}=2\beta_{A}$, and 
$\alpha_{max}=1/\left(\mathcal{Z}_{A}\mathcal{Z}_{B}\right)$.  
\label{fig_characterization}}
\end{figure}

In Fig.~\ref{fig_characterization} we plot the average work and efficiency for (a) and (c) initially uncorrelated qubits, (b) and (d) initially correlated qubits.  The blue dots give the results of 10 experiments for each energy gap ratio value shown in the figure. Each dot is obtained by  running the circuit Fig.~\ref{fig_transpiled} with 20000 shots, using the parameters given in Table~\ref{Tableangles}. Figures~\ref{fig_characterization} (a) and (b) demonstrate that the extractable work (negative work) under initially correlated qubits is larger (about an order of magnitude larger at its maximum) than  the one obtained in the absence of initial qubit correlations. In Fig.~\ref{fig_characterization} (c) and (d) we show the  efficiency obtained for the experimental  implementation of the thermal machine. This is in agreement with the criterion given below (see Eqs.~\eqref{criterion} and \eqref{criterionb}): efficiency for uncorrelated initial qubits remain below the standard Carnot limit, while initially correlated qubits may allow for a boost in efficiency, with values well above the Carnot limit.
%
\begin{figure*}
\begin{centering}
\includegraphics[scale=0.39]{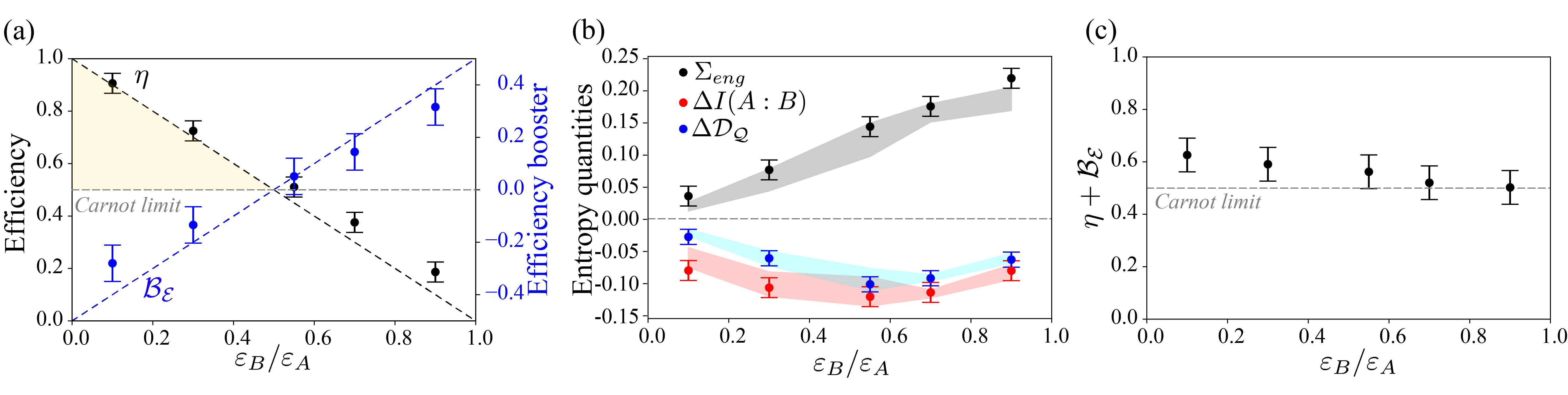}
\par\end{centering}
\caption{Experimental results for the correlated SWAP quantum heat engine implementation; figures of merit: (a) efficiency and booster, (b) entropy production, variation of mutual information and of quantum discord, and (c) test of the generalized second law limit for efficiency, which must satisfy $\eta + \mathcal{B_E} = \eta_{Carnot}=1/2$. The black and blue dashed lines in (a) correspond to the theoretical expectation for the correlated SWAP engine efficiency and efficiency booster, respectively.  The shaded areas in (b) denote the theoretical results but calculated with the actual initial experimental states (instead of the ideal ones) of the various experimental runs.
In all experiments, we set: 
$\beta_{B}=2\beta_{A}$,
 $\lambda=0.6$,
and 
$\alpha_{max}=1/\left(\mathcal{Z}_{A}\mathcal{Z}_{B}\right)$.
The error bars were estimated using the standard deviation of the measured data.
\label{fig3}}
\end{figure*}

\section{Boosting quantum engine efficiency by quantum correlations
}
\label{booster}

The mutual information $I(A:B) = S_A+ S_B-
S_{AB}$ gives a measure of the total correlations between systems $A$ and $B$, where 
$S_{i}=-\text{Tr}_i \left(\rho_i\ln\rho_i\right)$ is  the von Neumann entropy of state $\rho_i$.
We next derive an analytical expression for the SWAP engine efficiency.
Such efficiency involves an information-to-energy trade-off relation written
in terms of the single-cycle variation  of 
the mutual information between qubits $A$ and $B$, $\Delta I(A:B)=\Delta S_{A}+\Delta S_{B}$ (where 
$\Delta S_{i}=S(\rho_{i}^{f})-S(\rho_{i}^{0})$) and of the  entropy production cycle,  $\Sigma_{\text{eng}}=D[\rho_{A}^{f}||\rho_{A}^{0}]+D[\rho_{B}^{f}||\rho_{B}^{0}]$. Here, $D\left[\rho||\sigma\right]=\text{Tr}\left[\rho\left(\ln\rho-\ln\sigma\right)\right]$ is the
Kullback--Leibler divergence~\cite{Adesso2018}.
For the cycle, we arrive at the following  generalized efficiency (see Appendix~\ref{eta}): 
\begin{equation}
\eta=\eta_{Carnot}-\frac{\Sigma_{\text{eng}}+\Delta I(A:B)}{\beta_{B}\left\langle Q_A\right\rangle}.
\label{fluctuation}
\end{equation}
Equation~(\ref{fluctuation}) can be applied to all cycles based on bipartite working substance (of any dimension) when work and heat exchanges are performed in different strokes. Let us now introduce an efficiency booster quantifier
$\mathcal{B_E}$,
\begin{equation}
\mathcal{B_E}\equiv \frac{\Sigma_{\text{eng}}+\Delta I}{\beta_{B}\left\langle Q_A\right\rangle},
\label{lag}
\end{equation}
which tracks the direct competition between  entropy production 
and correlations consumption.
\begin{figure*}
\begin{centering}
\includegraphics[scale=0.45]{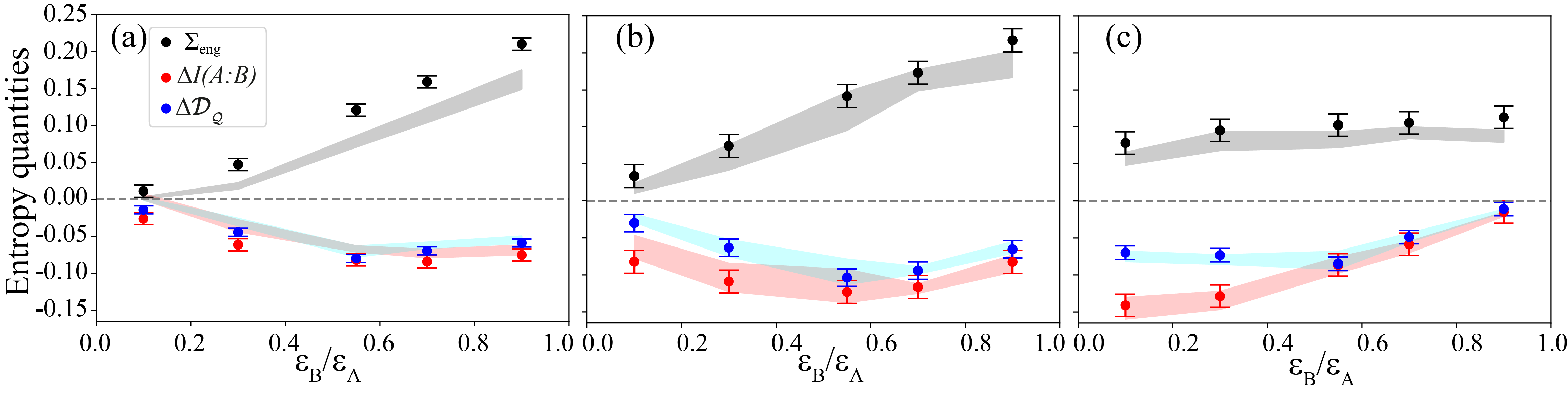}
\par\end{centering}
\caption{Entropy and correlation quantities 
during the thermodynamic cycle: (a) $\lambda=0.8$, (b) $\lambda=0.6$, and (c) $\lambda=0.2$. Experimental data in black, blue, and red correspond to the  entropy production, variation of  quantum discord and  variation of mutual information, respectively. In all runs in the quantum processor,  the temperature relation $\beta_{B}=2\beta_{A}$ and the correlation factor $\alpha_{max}=1/\left(\mathcal{Z}_{A}\mathcal{Z}_{B}\right)$.
 The shaded areas  denote the theoretical results calculated with the actual initial experimental states (instead of the ideal one), see e.g. Fig.~\ref{figtomo_sm}.
 The error bars were estimated using the standard deviation of the measured data.
\label{fig2_sm}}
\end{figure*}

\subsection{Engine efficiency criterion}
\label{eff}

Equation~\eqref{fluctuation} implies that there may exist  efficiencies above Carnot, $\eta>\eta_{Carnot}$, depending on the sign of $\Delta I$, with
the following engine efficiency criterion arising
\begin{eqnarray}
\mathcal{\mathcal{B_E}}<0, \;  \eta>\eta_{Carnot}, 
 \label{criterion}
   \\ 
\mathcal{\mathcal{B_E}}\geq 0, \; \eta\leq \eta_{Carnot}.
   \label{criterionb}
   \end{eqnarray}
Equation~(\ref{fluctuation}) can be seen as a quantum generalization of the second law efficiency and expression~\eqref{criterion} is the condition for performance over the classical limit to occur, and
indeed it is satisfied in
 Fig.~\ref{fig3}(a). This result arises since the variation of mutual information $\Delta I$ is always negative, and there is a trade-off with the always positive entropy production $\Sigma_{\text{eng}}$ (see Fig.~\ref{fig3}(b)). The
variation of $\Delta I=
\Delta\mathcal{D_Q} + \Delta\mathcal{C}$, where $\mathcal{D_Q}$ quantifies purely quantum correlations (here given by the quantum discord~\cite{
Ollivier2001}), and  $\mathcal{C}$ represents classical correlations, demonstrates that there is a consumption of quantum correlations during the thermodynamic cycle. For $\varepsilon_{B}/\varepsilon_{A}<1/2$, this makes  $\Sigma_{eng}<\lvert\Delta I(A:B)
\lvert$ and hence $\mathcal{\mathcal{B_E}}<0$,
which in turn implies
$\eta>\eta_{Carnot}$
in Eq.~\eqref{fluctuation}. 

For the calculation of the quantum correlations we have used the geometrical quantum discord $\mathcal{D_Q}$ for a two-qubit $X$ state~\citep{Dakic2010,Girolami2012}, which can be written in terms of the elements of the density matrix as:
\begin{equation}
   \mathcal{D_Q} = \left[k_1 - 2k_2 +4z^2 -max\left(k_1 -2k_2,2z^2\right) \right],
\end{equation}
where $k_1 =a^2 + b^2 + c^2 + d^2$ and $k_2 =ac + bd$. 

The emergence of condition~\eqref{criterion} depends on a proper choice of initially correlated states and of the  driving  Hamiltonian in the stroke 2. It only arises if  $\mathcal{B_E}<0$. Otherwise,
$\eta\leq \eta_{Carnot}$ (Eq.~\eqref{criterionb}); $\mathcal{B_E}\geq 0$
also describes engine operation for initially uncorrelated qubits (Fig.~\ref{figenergy}(a)):  entropy production is always greater or equal than  variation of mutual information. An experimental verification of the engine efficiency criterion Eqs.~\eqref{criterion} and \eqref{criterionb} (see Figs.~\ref{fig_characterization},~\ref{fig3}  and~\ref{fig2_sm}) is provided below.

\subsection{Experimental demonstration of performance boosting
in the correlated SWAP heat engine}
\label{exp}

In Fig.~\ref{fig3}, 
we plot  the quantum-processor-implementation results
for the efficiency, the entropy and the  generalized second law related quantities as a function of the energy gap ratio.
These quantifiers
have been obtained by using quantum state tomography (QST), as detailed below. 
The efficiency $\eta$ (black dots) and the efficiency booster $\mathcal{\mathcal{B_E}}$ (blue dots) are displayed in Fig.~\ref{fig3}(a). The plotted error bars were estimated as in  Fig.~\ref{figenergy}. 
The booster $\mathcal{\mathcal{B_E}}$ reaches negative values hence the
efficiencies go above the  Carnot limit, $\eta_{Carnot}=0.5$, in agreement with our theoretical findings (solid lines). 
In Fig.~\ref{fig3}(b), we plot the entropy production (black dots), the variation of mutual information (red dots), and of quantum discord (blue dots) obtained in the quantum processor implementation. 
For   $\varepsilon_{B}/\varepsilon_{A}<0.5$, we obtain   $\lvert\Delta I(A:B)
\lvert >\Sigma_{\text{eng}}$ and since $\Delta I$ is always negative,  $\mathcal{\mathcal{B_E}}<0$ and the SWAP engine efficiency surpasses the standard Carnot limit $\eta>\eta_{Carnot}$, which is in perfect agreement with Fig.~\ref{fig3}(a) and with the criterion Eq.~\eqref{criterion}. 

We make explicit the role of the measured correlations (both quantum and classical), by plotting  $\Delta I=
 \Delta\mathcal{D_Q} + \Delta\mathcal{C}$ and $\Delta\mathcal{D_Q}$, during the thermodynamic  cycle (see the two lower curves of Fig.~\ref{fig3}(b)). Here, the 
variation of 
quantum discord closely follows that of mutual information  and both are always negative. This means that while entropy production increases, both variations in classical and quantum correlations are consumed during the cycle. 
The larger  correlations consumption is of purely quantum origin, and come from the discord. These results have been further verified and plotted for $\Sigma_{\text{eng}}$,  $\Delta I$ and $
\Delta\mathcal{D_Q}$ in Fig.~\ref{fig2_sm}, for other values of $\lambda$.
The shaded areas in Fig.~\ref{fig3}(b) represent the theoretical results calculated with the actual initial states in the quantum processor (instead of the ideal one) of the various runs. The theoretical predictions are in very good agreement with the implementation results.

Figure~\ref{fig3}(c) shows the quantum-processor-implementation results for a verification of the generalized efficiency formula as a function of the qubits' energy-gap ratio. The black dots give the results for the sum of the engine efficiency and the booster $\eta+ \mathcal{\mathcal{B_E}}$, as averaged measurements following the qubits statistics from collected data from 20000 runs. We find that the generalized second law limit measured with the  \textsl{ibmq\_manila} quantum processor
is at most within two standard deviations from the analytical result  Eq.~\eqref{fluctuation}, 
thus verifying, with Fig.~\ref{fig3}(b), the quantum origin of  the working thermodynamical  principle for enhancing the efficiency of the correlated SWAP quantum heat engine.

\subsection{Entropy related quantities and quantum state tomography}
\label{tomo}

In Fig~\ref{fig2_sm} we plot the variation of entropic and correlation quantities as function of $\varepsilon_B/\varepsilon_A$, namely the variation of the engine entropy production  $\Sigma_{\text{eng}}$, of the mutual information $\Delta I$ and of the quantum discord $\mathcal{D_Q}$, for (a) $\lambda = 0.8$, (b) $\lambda = 0.6$, and (c) $\lambda = 0.2$. 
 Figure~\ref{fig2_sm} further demonstrates that the variation  of mutual information is mostly due to the consumption of quantum correlations between the qubits. It also confirms that for initially correlated systems $|\Delta I|$ may be greater than $\Sigma_{\text{eng}}$, leading to a correlation boosting of the engine efficiency.
\begin{figure*}
\begin{centering}
\includegraphics[scale=0.35]{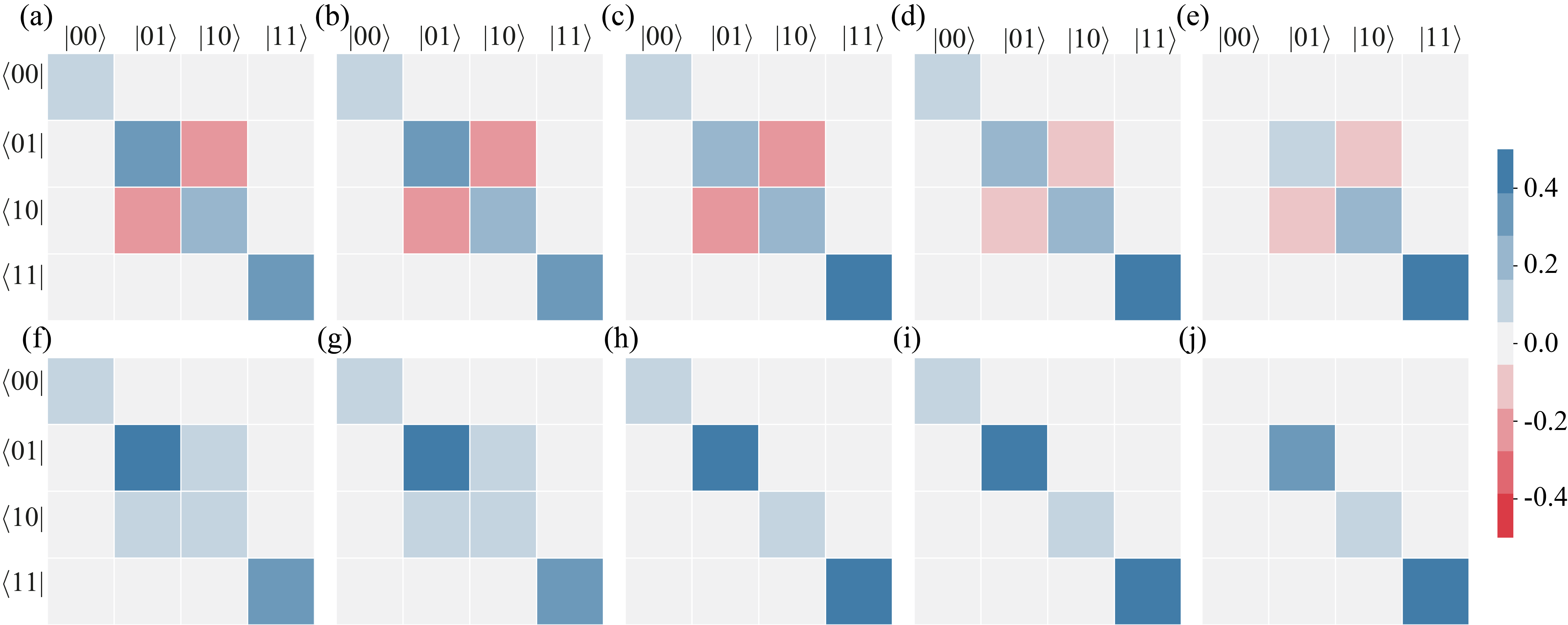}
\par\end{centering}
\caption{Quantum state tomography of the two-bit working substance for the initially correlated state (upper row) and corresponding final state (lower row) for $\lambda=0.6$ and for the following qubits energy gap ratios $\varepsilon_B/\varepsilon_A$: (a) and (f)  0.10; (b) and (g)  0.30; (c) and (h) 0.55; (d) and (i)  0.70; (e) and (j)  0.90. Data shown represent the result over 20000 shots for each circuit used in one of the QST implementations. Re$(\rho)$ 
 denotes   the real values of the density matrix (the corresponding  imaginary part entries are of the order of $10^{-3}$, not shown).}
\label{figtomo_sm}
\end{figure*}

For the quantum state tomography (QST)
implementation required in this analysis we have used a module in the qiskit-ignis library~\citep{qiskittomo}. 
For a  complete QST of a two-qubits state, nine  circuits are needed. The result for each circuit is averaged over 20000 shots and, additionally,
in order to average over the system fluctuations,
we repeat the process five times for the initial state and five times for the final one. In Figure~\ref{figtomo_sm}(a)-(e) we show the QST result for the initially correlated state $\rho^0_{AB}$, considering $\varepsilon_B/\varepsilon_A=0.10,0.30,0.55,0.70,0.90$, respectively. In the same way, in Fig.~\ref{figtomo_sm}(f)-(j) we give the QST result for the final state, with $\lambda = 0.6$ and $\varepsilon_B/\varepsilon_A=0.10,0.30,0.55,0.70,0.90$, respectively. Similar results were obtained for $\lambda = 0.2, 0.8$ (not shown). In the computational basis, the qubits density matrix can then be approximated as
\begin{equation}
\rho=\left(\begin{array}{cccc}
a & 0 & 0 & 0\\
0 & b & z & 0\\
0 & z & c & 0\\
0 & 0 & 0 & d
\label{xmatrix}
\end{array}\right).
\end{equation} 

\section{Summary}
\label{conc}

In summary, the limits posed by the second law of thermodynamics may
be affected by the presence of initial quantum correlations in the
working fluid of a thermal machine, leading to efficiency higher than
the Carnot standard limit and to a boost in the extractable work in each cycle. A criterion for the construction of such enhanced thermodynamic feature is given in terms of a trade-off between entropy production and quantum correlations consumption during the implemented thermal machine's cycle.
The design of thermal machines that use extra resources based on quantum correlations highlight the need for a revision of the standard thermodynamics limits. In this framework, the energetic cost of building initial correlations should not be included  in the efficiency definition. This is in line with the practice of not including costs related to the production of (hot) heat sources in classical internal-combustion engines (e.g., fuel production/refining, etc.). Our results with an IBM quantum processor clearly demonstrate that the effect we propose allows to obtain advantages in thermal tasks using available quantum technology.
 
\begin{acknowledgments}

M.H. and J.H.R acknowledge the  financial support from MinCiencias (Colombia) through a Postdoctoral Fellowship Award
(Grant No.~270-2021/CI 71295) and 
the Norwegian
Ministry of Education and Research, through the QTECNOS consortium (NORPART 2021-10436/CI 71331). R.M.S. also acknowledges CNPq, FAPESP (Brazil) and Ministry of Science and Technology (China), through the High-End Foreign Expert Program (Grant No.~G2021016021L). 
I.D.A. acknowledges the kind hospitality of the   Instituto de Física de S\~ao Carlos, University of  S\~ao Paulo, S\~ao Carlos (Brazil). We also  acknowledge
the use of  \textsl{ibmq\_manila} quantum processor
of IBM Quantum services~\citep{ibmq}.
\end{acknowledgments}

\appendix

\section{ABOUT ENERGY EXCHANGE, CORRELATIONS PRODUCTION AND EFFICIENCY}
\label{energyexchange}

Here we show the effect of taking into account the initial-state correlation production in the thermodynamic cycle. To do so, we start from the state in stroke 1.1~(Fig.~1(a)). Such a state,
$\tilde{\rho}_A^0 \otimes \tilde{\rho}_B^0$, explicitly reads
\begin{equation}
\left(\begin{array}{cccc}
p_+p_- & 0 & 0 &0\\
0 & p_+(1-p_-) & 0 & 0\\
0 & 0 & p_-(1-p_+) & 0\\
0 & 0 & 0 & (1-p_+)(1-p_-)
\end{array}\right).
\end{equation}
The populations of states $\tilde{\rho}_{A}^{0}$ and $\tilde{\rho}_{B}^{0}$ are chosen as $p_{\pm}=\left(p_{A}+p_{B}\pm\sqrt{(p_{B}-p_{A})^{2}+4\alpha^{2}}\right)/2$ to obtain the desired correlated state $\rho_{AB}^{0}=\rho_{A}^{0}\otimes\rho_{B}^{0}+\chi_{AB}$, where $p_{-}$ ($p_{+}$) denotes the ground state population  of $\tilde{\rho}_{A}^{0}$  ($\tilde{\rho}_{B}^{0}$). 
Hence,  $\rho_{AB}^{0}$ becomes
\begin{widetext}
\begin{equation}
\left(\begin{array}{cccc}
p_Ap_B - \alpha^2 & 0 & 0 &0\\
0 & p_A(1-p_B) +\alpha^2 & \alpha & 0\\
0 & \alpha & p_B(1-p_A) +\alpha^2 & 0\\
0 & 0 & 0 & (1-p_A)(1-p_B) -\alpha^2
\end{array}\right),
\end{equation}
\end{widetext}
and the von Neumann entropy of $\rho_A^0 \otimes \rho_B^0 + \chi_{AB}$ is the same than that of $\tilde{\rho}_A^0 \otimes \tilde{\rho}_B^0$.  In fact, using the expressions for $p_\pm$ 
and $p_i$,
we obtain the same eigenvalues for both density operators, 
$\rho_{AB}^{0}$
and $\tilde{\rho}_A^0 \otimes \tilde{\rho}_B^0$: 
\begin{equation} 
\begin{split}
\nu_{1}	&=	\frac{1}{2}\left[p_{A}+p_{B}-2p_{A}p_{B}+2\alpha^{2}-\sqrt{\left(p_{B}-p_{A}\right)^{2}+4\alpha^{2}}\right],\\
\nu_{2}	&=	\left(1-p_{A}\right)\left(1-p_{B}\right)-\alpha^{2},\\
\nu_{3}	&=	p_{A}p_{B}-\alpha^{2},\\
\nu_{4}	&=	\frac{1}{2}\left[p_{A}+p_{B}-2p_{A}p_{B}+2\alpha^{2}+\sqrt{\left(p_{B}-p_{A}\right)^{2}+4\alpha^{2}}\right].
\end{split}
\end{equation}

\begin{figure}
\begin{centering}
\includegraphics[scale=0.38]{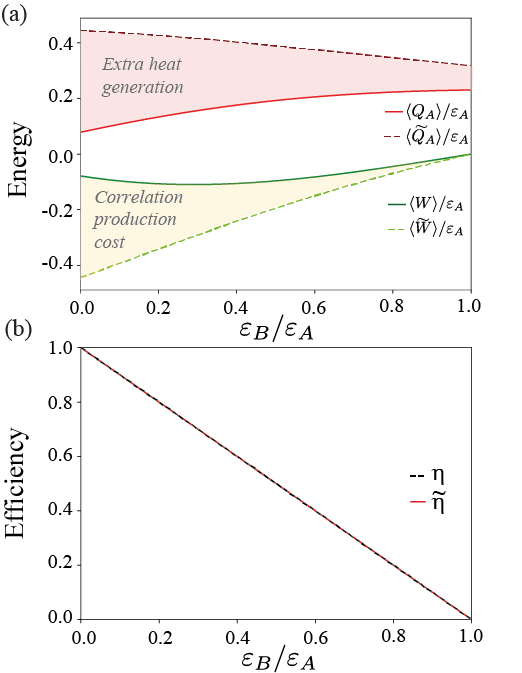}
\par\end{centering}
\caption{(a) Average work $\left\langle\tilde{W}\right\rangle/\varepsilon_A$, heat from the hot reservoir $\left\langle\tilde{Q}_{A}\right\rangle/\varepsilon_A$, and (b) efficiency $\tilde{\eta}$, including in  the cycle the initial-state correlation production. In the same graphs, we plot the  corresponding quantities but now  without including the initial-state correlation production  step: $\left\langle W\right\rangle/\varepsilon_A$, $\left\langle Q_{A}\right\rangle/\varepsilon_A$, and $\eta$. We set $\beta_{B}=2\beta_{A}$, 
$\lambda=0.6$ and $\alpha_{max}=1/\left(\mathcal{Z}_{A}\mathcal{Z}_{B}\right)$. \label{figs6}}
\end{figure}

From the initial state $\tilde{\rho}_A^0 \otimes \tilde{\rho}_B^0$, we calculate the heat  and work  as 
$\left\langle \tilde{Q}_{i}\right\rangle=-\text{Tr}\left[\left(\rho_{i}^{f}-\tilde{\rho}_{i}^{0}\right)\mathcal{H}_{i}\right]$ and $\left\langle\tilde{W} \right\rangle=\text{Tr}\left[\left(\rho_{AB}^{f}-\tilde{\rho}_{A}^{0}\otimes\tilde{\rho}_{B}^{0}\right)\mathcal{H}_{AB}\right]$, respectively.  In Fig.~\ref{figs6}(a), the dashed lines correspond to the heat generation ($\langle\tilde{Q}_A\rangle/\epsilon_A$, pink area) and work production ($\langle\tilde{W}_A\rangle/\epsilon_A$, yellow area) when considering $\tilde{\rho}_A^0\otimes\tilde{\rho}_B^0$ as the initial state. We  see that $\langle\tilde{Q}_A\rangle$ and $\langle\tilde{W}\rangle$ are greater than $\langle Q_A\rangle$ and $\langle W\rangle$ (in absolute value), but the cycle efficiency when considering the correlation production $\tilde{\eta} =\langle\tilde{W}\rangle/\langle\tilde{Q}_A\rangle$ corresponds to the same one computed from the correlated state $\eta=\langle W\rangle/\langle Q_A\rangle $, as it is  shown in Fig.~\ref{figs6}(b).  This is so because there is a compensation in the extra amounts obtained for the corresponding heat and work, as illustrated in  the shaded areas of Fig.~\ref{figs6}(a). These show  the extra heat generation (pink area) and the correlation production cost (yellow area) due to the generation of correlations in the initial state. 

\section{DEMONSTRATION OF THE GENERALIZED SECOND LAW LIMIT
}
\label{eta}

Here we 
give a demonstration 
for the efficiency result  Eq.~(\ref{fluctuation}). 
The entropy
production of the correlated SWAP quantum engine can be expressed as the sum of two relative entropies that reads
\begin{equation}
\begin{split}
D\left[\rho_{A}^{f}||\rho_{A}^{0}\right]+D\left[\rho_{B}^{f}||\rho_{B}^{0}\right]	 =& -\Delta S_{A}-\Delta S_{B}  \\
&+	\beta_{A}\text{Tr}\left[\left(\rho_{A}^{f} -\rho_{A}^{0}\right)\mathcal{H}_{A}\right] \\
&+\beta_{B}\text{Tr}\left[\left(\rho_{B}^{f}-\rho_{B}^{0}\right)\mathcal{H}_{B}\right],
\end{split}
\end{equation}
 where $\rho_{i}^{f}$ is the final out-of-equilibrium state for the qubit $i=A,B$. We simplify this equation by introducing $\left\langle Q_{i}\right\rangle=-\text{Tr}\left[\left(\rho_{i}^{f}-\rho_{i}^{0}\right)\mathcal{H}_{i}\right]$, hence
 \begin{equation}
\begin{split}
D\left[\rho_{A}^{f}||\rho_{A}^{0}\right]+D\left[\rho_{B}^{f}||\rho_{B}^{0}\right]=& -\Delta S_{A}-\Delta S_{B}\\
 &-\beta_{A}\left\langle Q_{A}\right\rangle-\beta_{B}\left\langle Q_{B}\right\rangle.
\end{split}
\end{equation}
We use the fact that $\Delta I(A:B)=\Delta S_{A}+\Delta S_{B}$ to  rewrite Eq.~(S4) as $ D\left[\rho_{A}^{f}||\rho_{A}^{0}\right]+D\left[\rho_{B}^{f}||\rho_{B}^{0}\right]=-\Delta I(A:B)-\beta_{A}\left\langle Q_{A}\right\rangle-\beta_{B}\left\langle Q_{B}\right\rangle$.
This is equivalent to
\begin{equation}
    \frac{D\left[\rho_{A}^{f}||\rho_{A}^{0}\right]+D\left[\rho_{B}^{f}||\rho_{B}^{0}\right]
    +\Delta I(A:B)}{\beta_{B}\left\langle Q_{A}\right\rangle}=-\frac{\beta_{A}}{\beta_{B}}-\frac{\left\langle Q_{B}\right\rangle}{\left\langle Q_{A}\right\rangle}.
\end{equation}
Energy conservation implies  that 
the average values for the heat and work,
$\left\langle Q_{A}\right\rangle
+\left\langle Q_{B}\right\rangle+\left\langle W\right\rangle
=0$. The extracted work  $\left\langle W_{ext}\right\rangle=-\left\langle W\right\rangle=\left\langle Q_{A}\right\rangle
+\left\langle Q_{B}\right\rangle$ and the quantum heat engine efficiency  $\eta=\left\langle W_{ext}\right\rangle/\left\langle Q_{A}\right\rangle$ reads

\begin{equation}
    \eta=1-\frac{\beta_{A}}{\beta_{B}}-\frac{D\left[\rho_{A}^{f}||\rho_{A}^{0}\right]+D\left[\rho_{B}^{f}||\rho_{B}^{0}\right]+\Delta I(A:B)}{\beta_{B}\left\langle Q_{A}\right\rangle},
    \label{gsl}
\end{equation}
where $1-\beta_{A}/\beta_{B}$ is the standard Carnot limit. Equation~\eqref{gsl}
defines
a generalized
second law limit for bipartite quantum engine in the presence of initial correlations. As discussed in the main text, the  efficiency booster
$\mathcal{B_E}\equiv (\Sigma_{\text{eng}}+\Delta I)/\beta_{B}\left\langle Q_{A}\right\rangle$ 
sets a criterion  (Eq.~\eqref{criterion}) for the  enhancement of the engine's  efficiency and extractable  work  (see Figs.~ \ref{fig1_sm},~\ref{fig_characterization} and~\ref{fig2_sm}).


\begin{thebibliography}{99}
\bibitem{Note_corr} At this point, we note note that non-classical correlations could be provided in various ways, e.g., by cooling an interacting many-body fluid to its ground/low energy state with related creation of (thermal) entanglement. This kind of non-classical correlations which are present in many-body thermal states could be used in further applications to obtain some quantum advantage in thermal protocols.
\end{thebibliography}
\end{document}